\documentclass{aa}  

\usepackage{graphicx,txfonts,natbib}
\bibpunct{(}{)}{;}{a}{}{,} 

\begin{document}
   \title{Optical properties of Y dwarfs observed with the Gran Telescopio Canarias
   \thanks{Based on observations made with the Gran Telescopio Canarias (GTC), installed in the Spanish 
Observatorio del Roque de los Muchachos of the Instituto de Astrof\'isica de Canarias, on the island of
 La Palma (programmes GTC27-13B and GTC49-14B -- P.I. N. Lodieu --  GTCMULTIPLE2B-21A 
and GTC133-21A -- P.I. E. L. Martín)}
   }


   \author{E.\ L.\ Mart\'in \inst{1,2}
            \and
      J.-Y.\ Zhang \inst{1,2}
               \and
      H.\ Lanchas \inst{3}
       \and
      N.\ Lodieu \inst{1,2}
       \and
      T.\ Shahbaz \inst{1,2}
        \and
     Ya.\ V.\ Pavlenko \inst{4,1}    
        }

   \institute{
       Instituto de Astrof\'isica de Canarias (IAC), Calle V\'ia L\'actea s/n, E-38200 La Laguna, Tenerife, Spain
       \and
       Departamento de Astrof\'isica, Universidad de La Laguna (ULL), E-38206 La Laguna, Tenerife, Spain
       \and
       Valencia International University, Spain
             \and
      Main Astronomical Observatory, Academy of Sciences of the Ukraine, 27 Zabolotnoho, Kyiv 03143, Ukraine
       }

   \date{Received July 7, 2023; accepted \today{}}

 
  \abstract
   {}
   {Our science goals are to characterise the optical properties of Y dwarfs 
and to study their consistency with theoretical models.}
   {A sample of five Y dwarfs  
   was observed with three optical and near-infrared instruments at the 
10.4\,m
Gran Telescopio Canarias. Deep near-infrared ($J$- or $H$-band) and 
multicolour
optical images ($z$-, $i$-, $r$-, $g$-, $u$-bands) of the five targets and 
a low-resolution far-red optical spectrum for one of the targets
were
obtained.   
   }
   {One of the Y dwarfs, WISE J173835.53+273258.9 (Y0), 
was
clearly 
detected in the optical ($z$- and $i$-bands)
and another, 
WISE J182831.08+265037.7 (Y2), 
was
detected only in the $z$-band. We measured the 
colours 
of our targets and found that the $z-J$ and $i-z$ 
colours
of the Y dwarfs are 
bluer than those of mid- and 
   late-T dwarfs. This optical 
blueing
has been predicted by models, but 
our data indicates that it is sharper and happens at 
temperatures 
about 150 K warmer 
than expected.  The culprit is the K I resonance doublet,
which 
weakens more abruptly 
in the T- to Y-type transition than expected. 
Moreover, we show that the alkali resonance 
lines (Cs I and K I) are weaker in Y dwarfs than in T dwarfs; the far-red optical spectrum of WISE J173835.53+273258.9 is 
similar to that of late-T dwarfs, but with stronger methane and water features; and 
we noted the appearance of new absorption features that we propose could be due to 
hydrogen sulphide. Last but not least,  in 2014, WISE J173835.53+273258.9 presented 
a bluer $i-z$ 
colour 
than in 2021
by a factor of 2.8 
   (significance of 2.5$\sigma$). Thanks to our deep optical images, we 
found 
that the 2014 $i$-band spectrum was contaminated by a galaxy bluer than the Y dwarf.}
{The optical properties of Y dwarfs presented here 
pose
new challenges to the 
modelling of grain sedimentation in extremely cool objects. The weakening of the 
very broad K I resonance doublet due to condensation in dust grains is more abrupt 
than theoretically anticipated. Consequently, the observed 
blueing
of 
the 
$z-J$ and 
$i-z$ 
colours
of Y dwarfs with respect to T dwarfs is more pronounced than predicted 
by models and could boost the potential of upcoming deep large-area optical 
surveys 
regarding their ability to 
detect
 extremely cool objects.}

\authorrunning{Mart\'in et al.  
  }
  \titlerunning{Optical properties of Y dwarfs observed with the GTC}

   \maketitle
%

%
%
\section{Introduction}
\label{intro}
%

With the discovery of an object with an effective temperature (T$_{\rm eff}$) below 
700 K in the Canada France Brown Dwarf survey, it was proposed that a new 
spectral class, dubbed Y, was needed to classify objects cooler than T \citep{delorme08a}. 
Spectral subclasses
for Y dwarfs 
began
to be defined after the discovery of 
seven objects with the {\it Wide-field Infrared Survey Explorer} (WISE) by 
\citet{cushing11},
who 
proposed that WISE J173835.53+273258.9 (hereafter W1738) 
be considered as
the Y0 spectral standard.  
The effective temperatures 
of the Y dwarfs reported by those authors 
range from 450 K to 300 K.

\begin{table*}
\tiny
\footnotesize
 \centering
 \caption[]{Sample of Y dwarfs with astrometric and photometric data from the literature.
}
 \begin{tabular}{@{\hspace{0mm}}l @{\hspace{1mm}}c @{\hspace{1mm}}c @{\hspace{1mm}}c @{\hspace{1mm}}c 
 @{\hspace{1mm}}c @{\hspace{1mm}}c 
 @{\hspace{1mm}}c @{\hspace{1mm}}c @{\hspace{1mm}}c @{\hspace{1mm}}c 
 @{\hspace{1mm}}c @{\hspace{1mm}}c@{\hspace{0mm}}}
 \hline
 \hline
ID &    R.A.     &   Dec. & epoch &   $i_{AB}$ & $z_{AB}$ & $J_{MKO}$ & $H_{MKO}$ & $\mu_\alpha$ & $\mu_\delta$ & d & T$_{eff}$ \cr
 \hline
 
   & hh:mm:ss.ss & ${^\circ}$:$'$:$''$ &MJD & mag & mag & mag & mag & arcsec/yr & arcsec/yr & pc & K  \cr
   
 \hline
 W1405  & 14:05:16.95 & 55:34:22.43 & 57249.47 &         & $>$23.85       & 21.06$\pm$0.06 & 21.41$\pm$0.08 & $-$2.33 & 0.23 & 6.3 & 411  \cr
 W1738  & 17:38:35.66 & 27:32:57.13 &57094.49 & $>$25.30 & 22.80$\pm$0.09 & 19.58$\pm$0.04 & 20.34$\pm$0.08 & 0.34 & $-$0.34 & 7.6 & 450 \cr
W1828  & 18:28:31.46 & 26:50:38.65 &57094.09 &  & $>$24.46 & 23.48$\pm$0.30 & 22.73$\pm$0.13  & 1.02 & 0.17 & 10.0 & 406 \cr
W1935  & 19:35:18.64 & $-$15:46:20.51 &57939.58 & & & 23.93$\pm$0.33 & & 0.29 & $-$0.04 & 14.4 & 367 \cr
W2354  & 23:54:03.00 & 02:40:11.63 &57860.05 & &  & 22.72 $\pm$0.13 & 22.53$\pm$0.28 & 0.50 & $-$0.40 & 7.7 & 388 \cr
 \hline
\label{table_sample}
\end{tabular}
\end{table*}

It has been predicted that the $i-z$ 
colour of
ultracool dwarfs with T$_{\rm eff}$ 
in the range 
2000--400 K
is extremely sensitive to grain sedimentation in 
the atmosphere \citep{marley02} and,
 to a lesser extent, is also
 sensitive to metallicity
 \citep{leggett12a}. Recent work indicates that the combination of optical and 
infrared 
colours
can provide sensitive diagnostics to break the degeneracy between 
metallicity and temperature in ultracool dwarfs \citep{zhang23}.

In view of the upcoming new generation of 
deep large-area optical
surveys, such 
as those to be carried out with \textit{Euclid} and the Vera Rubin Legacy Survey of Space 
and Time, it is a timely moment to improve the characterisation of benchmark Y dwarfs at 
optical wavelengths in order to improve 
our 
understanding of these elusive objects and guide 
the search for new ones using 
the
resources
just mentioned.
So far, just four Y dwarfs have 
been detected in the optical ($z$-band) with the 10.4\,m Gran Telescopio Canarias 
(GTC) by \citet{lodieu13a}, and only one far-red optical spectrum has been 
presented as 
a
Y dwarf, which was obtained with Gemini Multi-Object Spectrograph (GMOS) at the 8.1\,m Gemini North 
telescope \citep{leggett13}. 

In this 
paper,
we present a study of 
the 
optical properties of Y dwarfs based on new GTC observations.  
In Section~\ref{observations}, our sample of 
five Y dwarfs is described, and the imaging and spectroscopic observations using the GTC are presented.
Section~\ref{results} 
provides our photometric and spectroscopic results and 
puts 
them in context with other 
ultracool dwarfs. Section~\ref{discussion}
includes a discussion of our results and critical comparisons with theoretical models.
 Section~\ref{conclusions} summarises our main conclusions.

\section{Observations}
\label{observations}

A sample of five Y dwarfs was selected for observations with the GTC. The list of targets and their 
basic parameters are presented in Table 1.  Astrometric data, {\bf trigonometric} 
distances,
 and temperature 
estimates
 were taken from table 5 of \citet{Kirkpatrick2021}. 
The $i$-band
 and $z$-band photometry were obtained by \citet{lodieu13a} using GTC/OSIRIS with
 Sloan filters under 
the 
AB system,
and the near-infrared photometry under the MKO system 
was
taken from table 10 of \citet{leggett21}.

\begin{table}
\tiny
\footnotesize
\centering
 \caption[]{Total on-source exposure times
of the observations with the GTC/HiPERCAM and GTC/EMIR.}
 \begin{tabular}{@{\hspace{0mm}}l 
 @{\hspace{1.5mm}}c 
 @{\hspace{1.5mm}}c 
 @{\hspace{1.5mm}}c
 @{\hspace{1.5mm}}c
 @{\hspace{1.5mm}}c @{\hspace{1.5mm}}c
 @{\hspace{1.5mm}}c @{\hspace{1.5mm}}c
 @{\hspace{0mm}}}
 \hline
 \hline
ID   & $u$  &  $g$ & $r$ & $i$ & $z$ &   $J$     &  $H$   & R300R \cr
 \hline
   & s & s & s & s & s & s & s & s \cr
 \hline
 W1405 & 3858 & 3858 & 3858 & 3858 & 3858 &     & 2000 &   \cr
W1738  & 7488 & 7488 & 7488 & 7488 & 7488 & 360 &      & 32400  \cr
W1828  & 7716 & 7716 & 7716 & 7716 & 7716 &     & 2400 &   \cr
W1935  & 5787 & 5787 & 5787 & 5787 & 5787 &     & 2400 &    \cr
W2354  & 7716 & 7716 & 7716 & 7716 & 7716 & 2100 &     & \cr
 \hline
\label{texpsample}
\end{tabular}
\end{table}
\subsection{Direct imaging and photometry}
\label{phot}

\begin{figure}
 \centering  \includegraphics[width=9cm, angle=0]{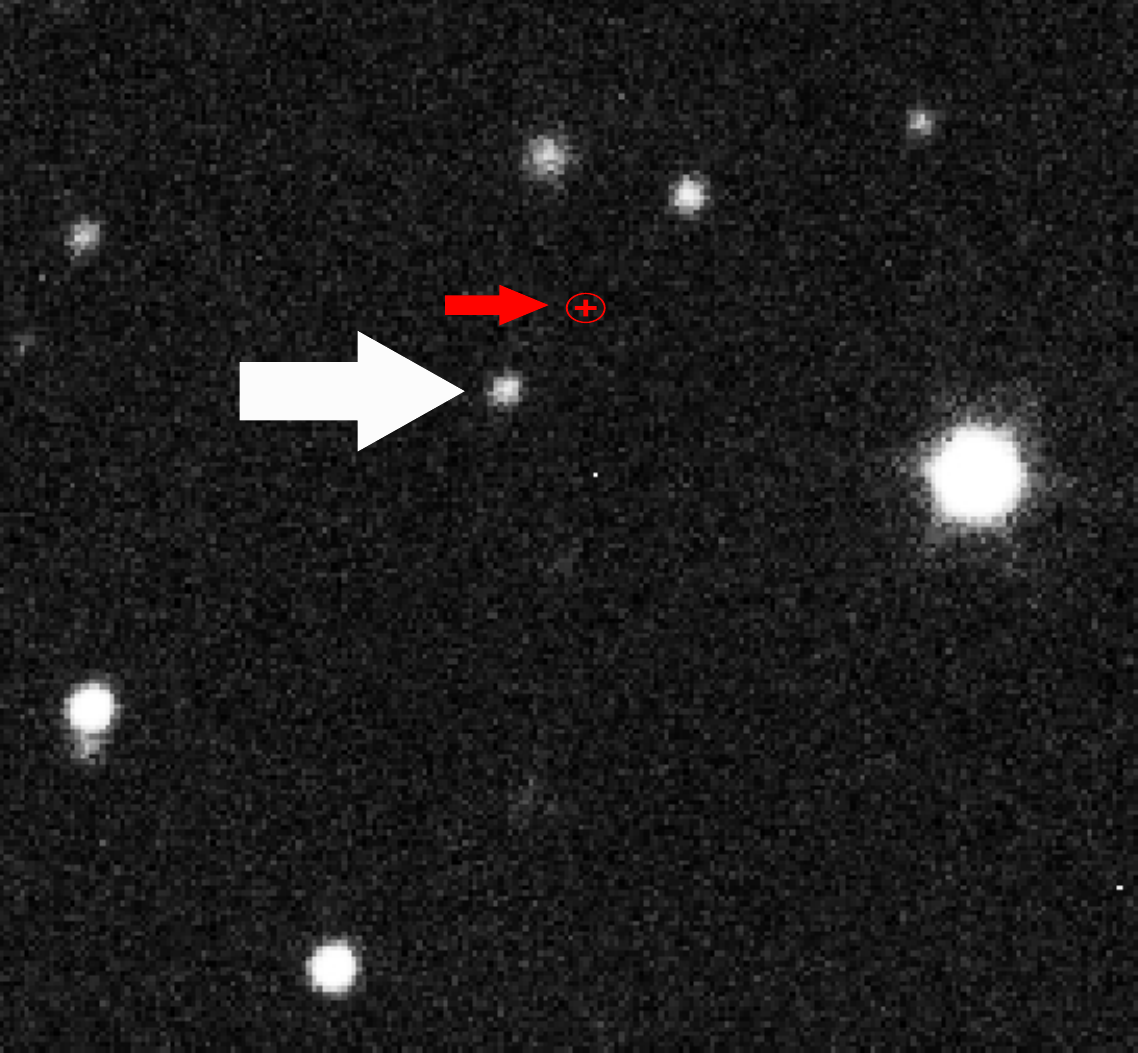}
  \caption{ GTC/EMIR $J$-band image of W1738. North is up and east is to
 the left. The size of the image is 43 arcsec on each side. The position 
of W1738  listed in \citet{lodieu13a} is marked with a red circle.  W1738
 was found in our EMIR image at the expected position according to the 
published proper motion.}
 \label{emir_image}
\end{figure}
\begin{figure}
 \centering
\includegraphics
[width=9cm, angle=0]{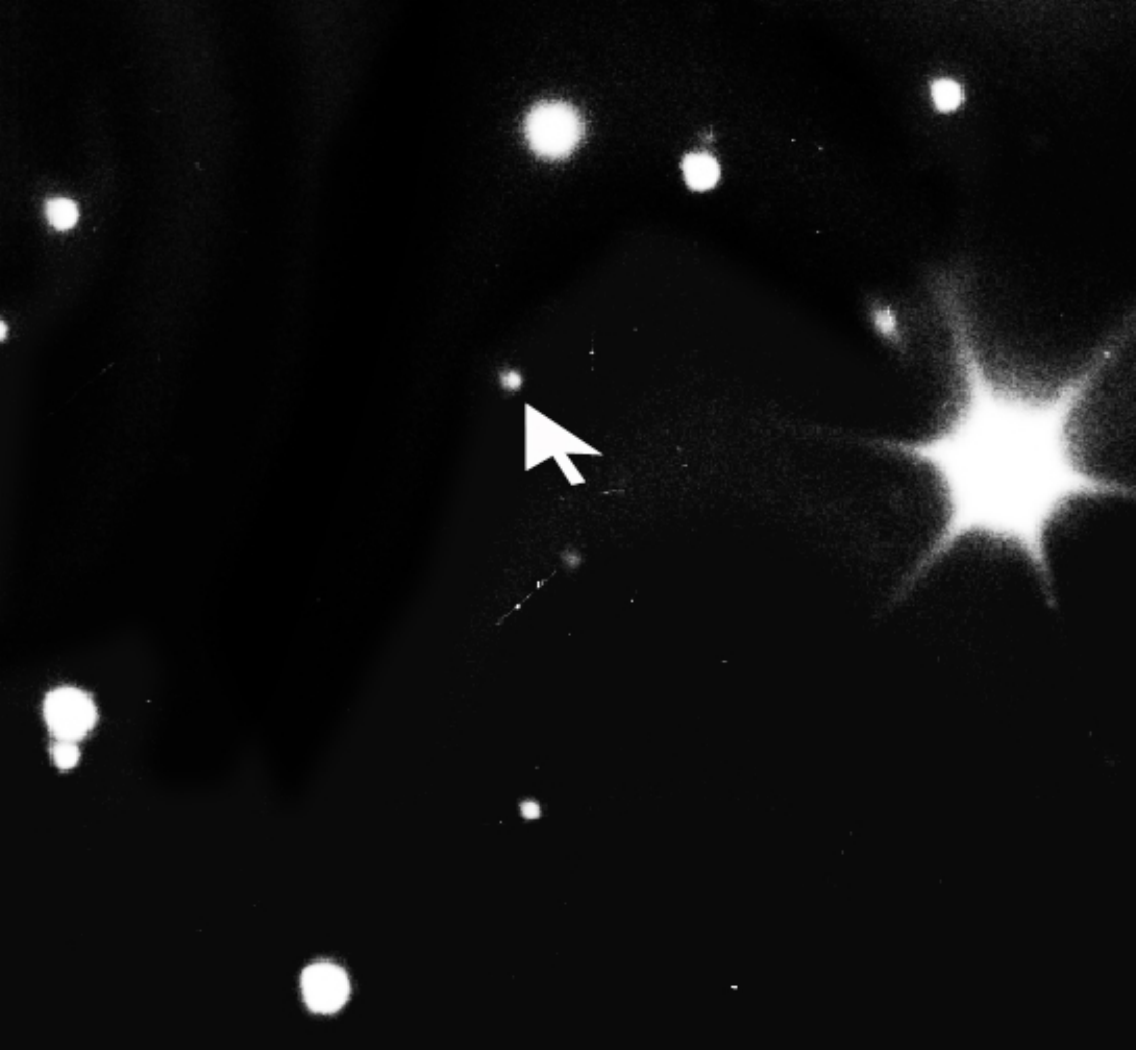}
\includegraphics[width=9cm, angle=0]
{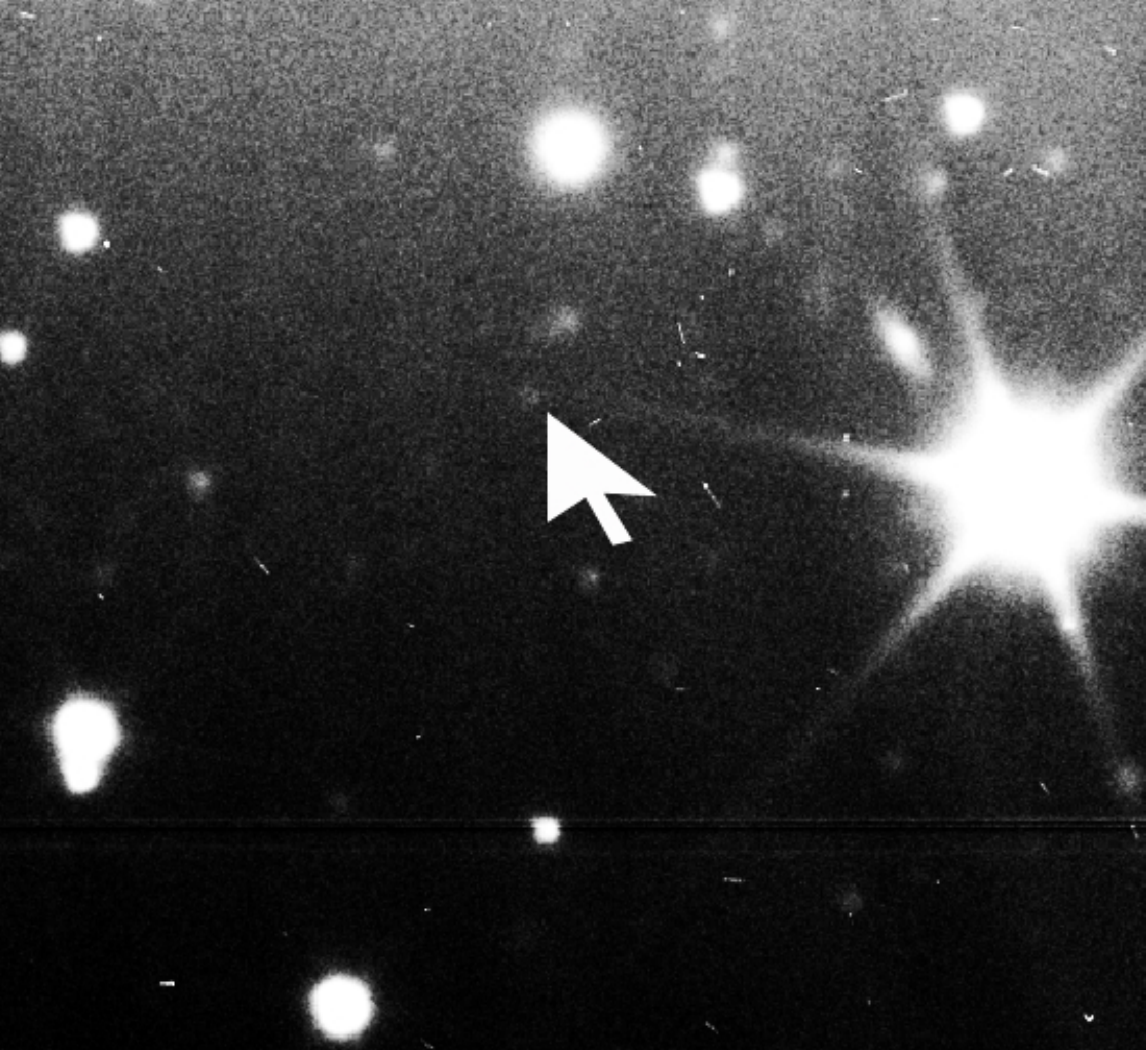} 
  \caption{ GTC/HiPERCAM images of W1738 in the $z$ filter (upper image) 
and the $i$ filter (lower image). The position of the Y dwarf is marked
 with an arrow. 
The
orientation and size of the images are the same as in 
Figure 1. W1738 is clearly detected in these two images  but not in the other
 three HiPERCAM filters ($r$, $g$, and $u$).
  We note the presence of a galaxy just north of W1738 that is seen in the $i$ 
filter but not in the $z$ filter. W1738 was much closer to that galaxy in 
2014 than in 2021, and the $i$-band long-slit spectrum obtained with OSIRIS
 in 2014 suffered from contamination by this galaxy.}
 \label{hipercam_image}
\end{figure}

\begin{figure}
 \centering  \includegraphics[width=9cm, angle=0]{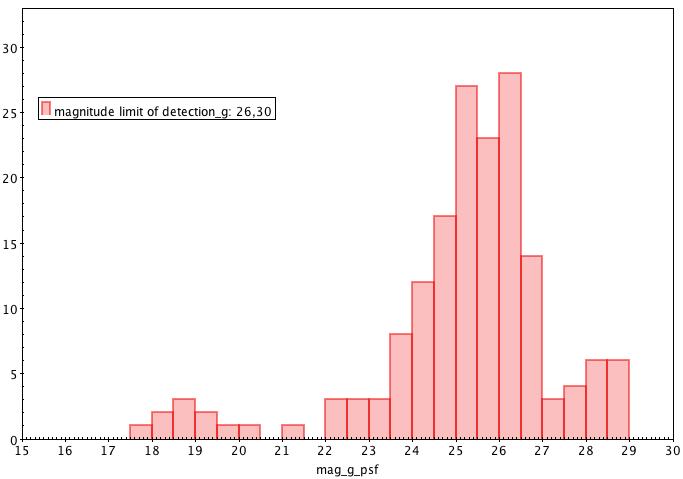}
  \caption{Histogram of GTC/HiPERCAM point source detections with PSF photometry 
in the $g$-band in the FOV of W1738.}
 \label{hisplot}
\end{figure}

\begin{figure}
 \centering
 \includegraphics[width=9cm, angle=0]{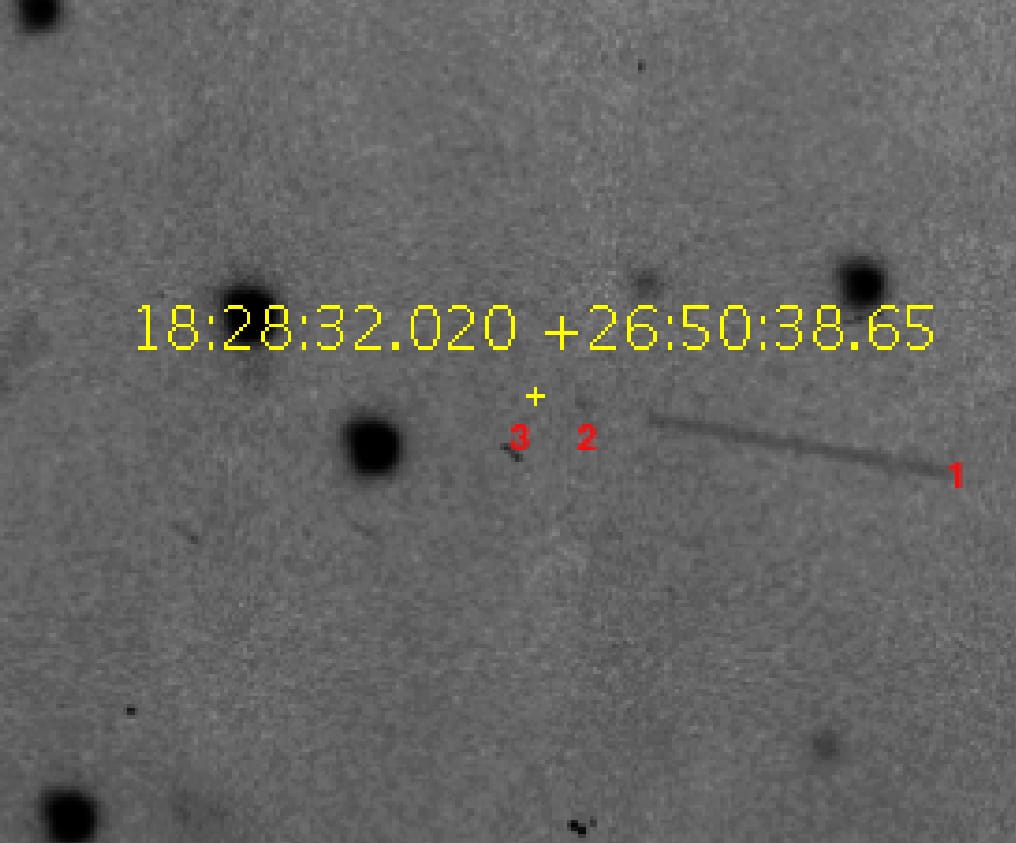}
\caption{GTC/HiPERCAM detection of W1828 in the $z$-band. The position 
labelled 1 was obtained with 
\textit{HST} 
in May 2011 
\citep{Beichman2013}. The dark segment indicates the proper motion vector. 
The position labelled 3 was obtained with 
\textit{JWST} in July 2022 
\citep{Defurio2023}.
The position labelled 2 is our detection with GTC in August 2021.}
 \label{hipercam1828}
\end{figure}

Direct images of the sample of five Y dwarfs were obtained with two different 
instruments at the GTC in service mode for the
programme  
GTCMULTIPLE2B-21A 
(PI E.L. Martín). The total amount of telescope time,
 including overheads, 
was 18 hours. Total on-source exposure times per filter are listed in Table 2.
 The data were collected from 
2021 May 11
to 
2021 August 29.

The Espectrógrafo Multiobjeto Infra-Rojo (EMIR) is a common user near-infrared imager and 
spectrograph located at the 
Nasmyth 
focus of the GTC \citep{Garzon2022}. It 
has a plate scale of 0.1945 arcsec/pix and a field of view (FOV) of 6.67$\times$6.67 
square arcmin in the imaging mode. The Y dwarfs were observed using the EMIR imaging
mode in the $J$-band or $H$-band filters. Standard data reduction was carried 
out using the PyEMIR pipeline. The final EMIR image of W1738 is shown in 
Figure~\ref{emir_image}.

At the time of our observations, HiPERCAM was a visiting instrument at the GTC 
\citep{Dhillon2021} with a plate scale of 0.081 arcsec/pix and an FOV 
of 2.8$\times$3.1 square arcmin. 
Its capable of obtaining simultaneous images in five different filters across the optical range 
($u,g,r,i,z$) using dichroics and five different detectors. Standard data 
reduction was carried out using the HiPERCAM pipeline. The final HiPERCAM images 
of W1738 in the $i$ and $z$ filters are shown in Figure~\ref{hipercam_image}.

Photometric measurements 
of
the images were performed 
in
the  Image Reduction and Analysis Facility (IRAF) environment 
\citep{tody86,tody93}.\footnote{IRAF
is distributed by the National Optical Astronomy Observatory, which is operated by 
the Association of 
Universities for Research in Astronomy (AURA) under cooperative agreement with the
 National Science Foundation.} 
For HiPERCAM images, aperture photometry was carried out with the \texttt{phot} task 
using a fixed aperture of 4 pixels and a variable aperture of 2.5 times the full width half maximum (FWHM).
The FWHM values measured in the HiPERCAM W1738 images ranged from 7.9 to 8.4 
pixels (0.64--0.68 arcsec). The sky contribution was measured in an annulus 10 
pixels away from the source and 10 pixels wide in radius. The point spread function (PSF) photometry was computed
with the \texttt{allstar} task. The results obtained with the PSF photometry had slightly smaller 
error bars than (but consistent with) those obtained with aperture photometry. Hence, 
we adopted the PSF photometric results. Five reference stars in the FOV 
of W1738 were used for establishing the zero point. These stars have photometric data 
in the catalogue released by the Panoramic Survey Telescope
and Rapid Response System \citep[Pan-STARRS;][]{chambers16a}.  For EMIR images, we also 
performed PSF photometry, and the zero point was established using five or six reference 
stars with photometry measured by the Two Micron All Sky Survey \citep[2MASS;][]{cutri03}. 

A histogram showing the number of sources detected with the IRAF \texttt{daofind} routine in the 
$g$-band is shown in Figure~\ref{hisplot}. From this histogram, we estimated the magnitude 
limit of detection of W1738 in the $g$-band. A similar procedure was followed for W1738 in
 the $r$-band and for W1828 in the $g$-band and the $i$-band. The results obtained 
are either consistent or more conservative than the 5$\sigma$ limiting magnitudes published on 
the HiPERCAM GTC website\footnote{  http://www.gtc.iac.es/instruments/hipercam/hipercam.php}. For 
the rest of the targets and filters, upper limits for the non-detections were derived 
using the sensitivities provided in the HiPERCAM GTC website. The adopted photometric 
measurements and upper limits are listed in Table \ref{mag}.

The detection of W1828 in the $z$-band is shown in Figure~\ref{hipercam1828}. 
The value of $z$-band photometry for this object measured in our GTC/HiPERCAM image is 
brighter than the photometry obtained from \textit{JWST} in the F090W filter ($26.16 \pm 0.20$ mag) 
by  \citet{Defurio2023}. This can be explained by the different passbands of these two 
filters. The F090W filter has a sharp cut-off at 1.00 micron, while the $z$ filter of 
HiPERCAM still has significant transmission up to about 1.06 microns. The very steep 
optical spectrum of Y dwarfs implies that there is a significant contribution to the 
photometric magnitude in the spectral range between 1.00 and 1.06 microns. 

\begin{table}
\tiny
\footnotesize
 \caption[]{GTC apparent magnitudes and upper limits for non-detections of Y dwarfs.
}
 \begin{tabular}{@{\hspace{0mm}}l 
 @{\hspace{1mm}}c 
 @{\hspace{1mm}}c 
 @{\hspace{1mm}}c
 @{\hspace{1mm}}c
 @{\hspace{1mm}}c @{\hspace{1mm}}c
@{\hspace{1mm}}c
 @{\hspace{0mm}}}
 \hline
 \hline
ID & $u_{AB}$ & $g_{AB}$ & $r_{AB}$   &  $i_{AB}$ & $z_{AB}$  &   $J_{2MASS}$     &  $H_{2MASS}$    \cr
 \hline
   & mag & mag & mag & mag & mag & mag & mag  \cr
 \hline
 W1405   & $>$25.5 & $>$26.0 & $>$25.5 & $>$25.5 & $>$25.0 & & $>$20.2     \cr
W1738   & $>$25.5 & $>$26.3 & $>$25.1 &  25.59$\pm$0.16 & 22.63$\pm$0.12 & 19.48$\pm$0.04 &    \cr
W1828  & $>$25.5 & $>$26.6 & $>$25.5 & $>$25.2 & 25.20$\pm$0.20  &       & $>$19.8     \cr
W1935 & $>$25.5 & $>$26.0 & $>$26.0 & $>$26.0 & $>$25.5 &                & $>$21.0      \cr
W2354 & $>$25.5 & $>$26.5 & $>$26.0 & $>$26.0 & $>$25.5 & 23.11$\pm$0.38 &      \cr
 \hline
\label{mag}
\end{tabular}
\end{table}
\subsection{Spectroscopy}
\label{spec}

Far-red 
optical spectra of W1738 were obtained with the Optical System 
for Imaging and low Resolution Integrated Spectroscopy
\citep[OSIRIS;][]{cepa00} instrument at the GTC over two semesters in 2014 
(GTC49-14B; PI N. Lodieu) and 2021 
(GTC133-21A; PI E.L. Mart\'in). The OSIRIS detector consists of two 
2048$\times$4096 Marconi CCD42-82 with 
an 8 arcsec gap between them and 
operates at optical wavelengths from 365 to 1000 nm. The unvignetted 
instrument FOV is about 
7$\times$7 arcmin$^2$
 with a pixel scale of 0.125 arcsec. 

The GTC49-14B observations consisted of six observing blocks, each of them made of four 1800 s long exposures  shifted along the slit in a AABB pattern. They were obtained on three consecutive nights in visitor mode: 2014 August 23, 24, and 25. The 
GTC133-21A data were collected in service mode from 
2021 June 3 to 2021 July 7.
These observations were divided into three observing blocks that employed two 
on-source integrations of 1800 s each with two repeats shifted along the slit by 
15 arcsec (AABB pattern). Each block was observed on
a different night. The nights were dark and the seeing was better than 0.9 arcsec. The FOV was identified and the target was centered on the slit using two exposures of 60 s in the Sloan $z$-band filter for each observing block. 
All the spectra were 
acquired in the parallactic angle. 
The total on-source exposure time of the 2014 plus the 2021 observations is provided 
in Table~\ref{texpsample}.

 {For both the 2014 and the 2021 observations, the R300R grating and a CCD binning 
of 2$\times$2 were used. In 2014, the slit width was set 
at
1.5 arcsec, while in 
2021 it was set 
at
1.0 arcsec.  These configurations yielded a spectral resolving 
power of $R \approx 190$ and 280 at the wavelength of 900 nm.}
Bias, flats, and arc lamps were obtained during the afternoon or morning of the respective nights. 
We also observed the spectro-photometric standard star, Ross\,640 
\citep[DZA5.5;][]{greenstein67,harrington80,Wesemael93}, with the same OSIRIS configuration 
and the $z$ filter to correct for the second-order contamination and calibrate our target 
as far to the red as possible. 

 The OSIRIS spectroscopic data were reduced with the 
\textit{PypeIt} 
pipeline
\citep{pypeit:zenodo,pypeit:joss_pub}. The \textit{PypeIt} pipeline automatically calibrates 
the frame and subtracts the sky, extracts the object spectra, 
and
performs the wavelength 
and flux calibrations in a standard manner.
The sensitivity function for the instrumental response correction was generated from 
the standard star spectra observed in the morning 
of 2012 July 8
with the same 
grism but with a slitwidth of 2.5 arcsec, with and without 
the 
$z$-band filter. We 
spliced the part before $9000$ \AA\ of the non $z$-filtered standard spectrum and 
the part after $9000$ \AA\ of the $z$-filtered standard spectrum to obtain the 
standard spectrum covering the whole wavelength range without second-order contamination. 
We note that the spectrum of the target W1738 is red enough to be free of second-order 
contamination; 
hence no splicing was needed. The pipeline co-added all the spectra.
We asked the pipeline to mask the water absorption region from $9150$ \AA\ to $9900$ \AA\ 
and to fit a low-order polynomial to the co-added spectrum. The
pipeline 
then 
used the 
Mauna Kea telluric grids to correct the spectra for the telluric contribution.

\section{Results}
\label{results}
\subsection{Photometric results}
\begin{figure}
 \centering
 \includegraphics[width=9cm, angle=0]{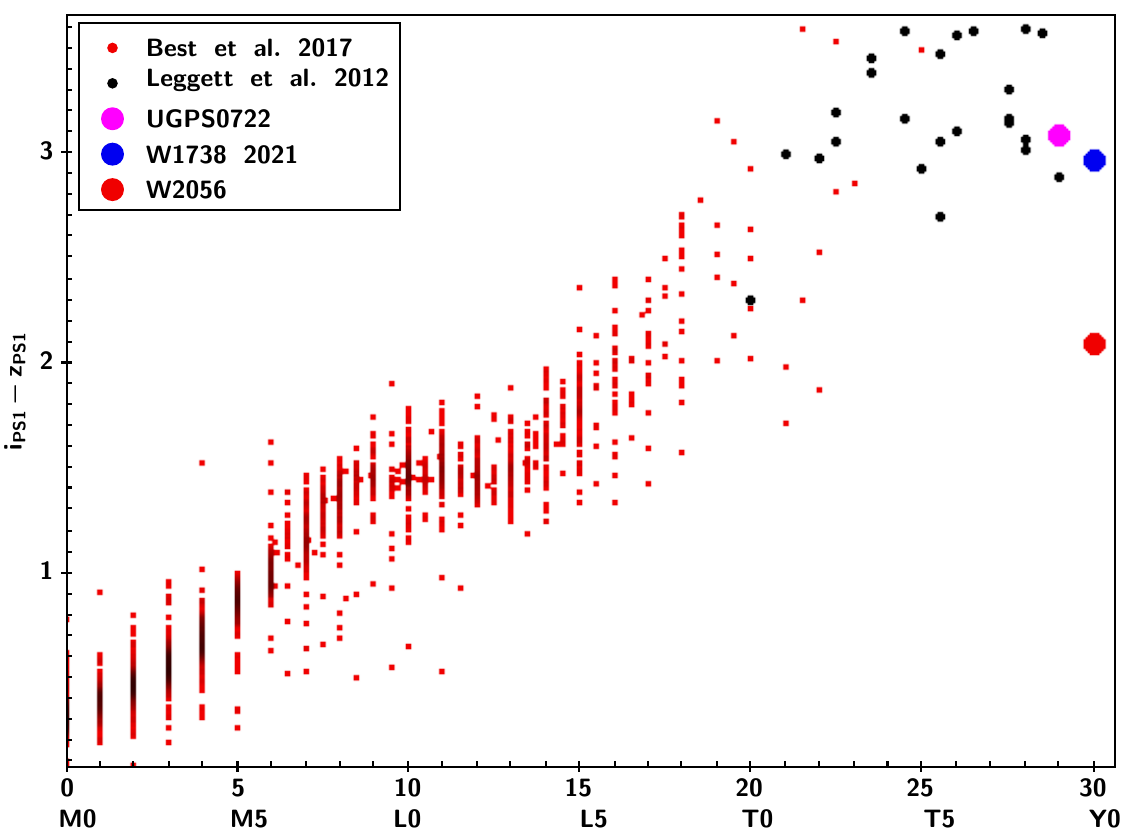}
 \includegraphics[width=9cm, angle=0]{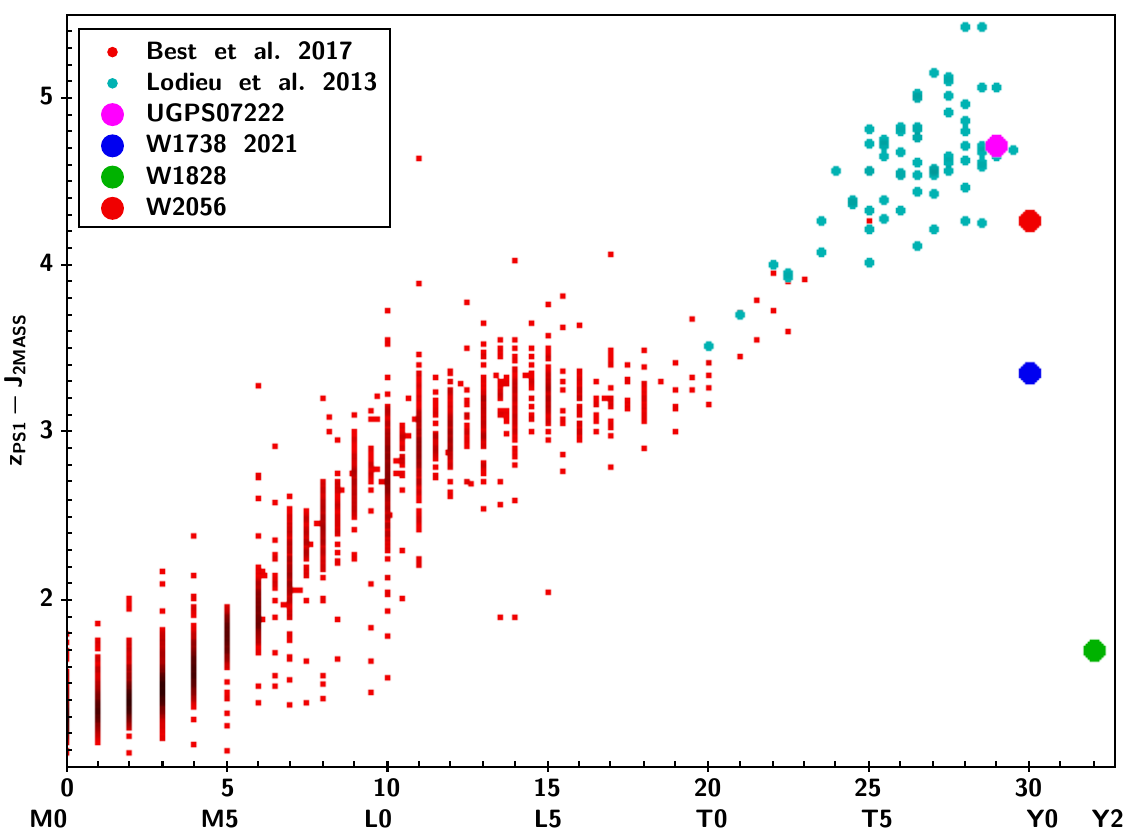}
 \caption{Colours with respect to spectral type for cool and ultracool dwarfs. The adopted 
spectral type of W1738 is Y0 \citep{cushing11} and that of W1828 is Y2 \citep{Cushing2021}. 
Upper panel:  $i-z$ 
colour 
of W1738 obtained by GTC/HiPERCAM photometry in 2021 compared 
with dwarfs with photometry from \citet{best17a, leggett12a}, a T9 dwarf UGPS0722 with 
photometry from \citet{leggett13}, and a Y0 dwarf W2056 with systhesised $i-z$ 
colour using 
a 
spectrum from \citet{leggett13}. Lower panel: 
 $z-J$ 
colour 
of W1738 compared with dwarfs with photometry from \citet{lodieu13a}, 
\citet{best17a}, the $z$ magnitude of W2056 \citep{lodieu13a}, the $J$ magnitude of W1828, 
W2056 \citep{leggett13}, and UGPS0722 \citep{tinney18}. The photometry systems were under 
the PS1 AB system and 
the
2MASS system. The photometric errors are smaller than the data points.}
 \label{colorplot}
\end{figure}

Our GTC/HiPERCAM photometry for the two detected Y dwarfs was put 
into the context 
of cool and ultracool dwarfs from the literature. The $i-z$ and $z-J$ 
colours 
versus spectral type diagrams are shown in Figure~\ref{colorplot}.  The $z$ magnitudes 
were corrected for 
the
 PS1 AB system according to \citet{zhang23}. Specifically, the 
corrections for W1738, W2056, and UGPS0722 were done using their spectra directly.  
According to \citet{leggett06}, the correction of 
the
$J$ magnitude from 
the
MKO system to 
the
2MASS 
system is not significant in our plots. In both 
colours, 
the Y dwarfs tend to be similar 
to 
or bluer than the T dwarfs. This is particularly noticeable in the trend of the
$z-J$ 
colour 
with respect to spectral type. The $z-J$ 
colour 
of W1828 is much bluer than that of 
T dwarfs. We note that the $i-z$ 
colour 
of W1738 in Figure~\ref{colorplot} was obtained 
photometrically in 2021, and we noticed a variability 
in the 
$i-z$ 
colour 
using multiple-epoch 
spectroscopy, which we discuss in Section~\ref{discussion} and Figure~\ref{specplot_epoch}.

\subsection{Spectroscopic results}
\begin{figure}
 \centering
 \includegraphics[width=9cm, angle=0]{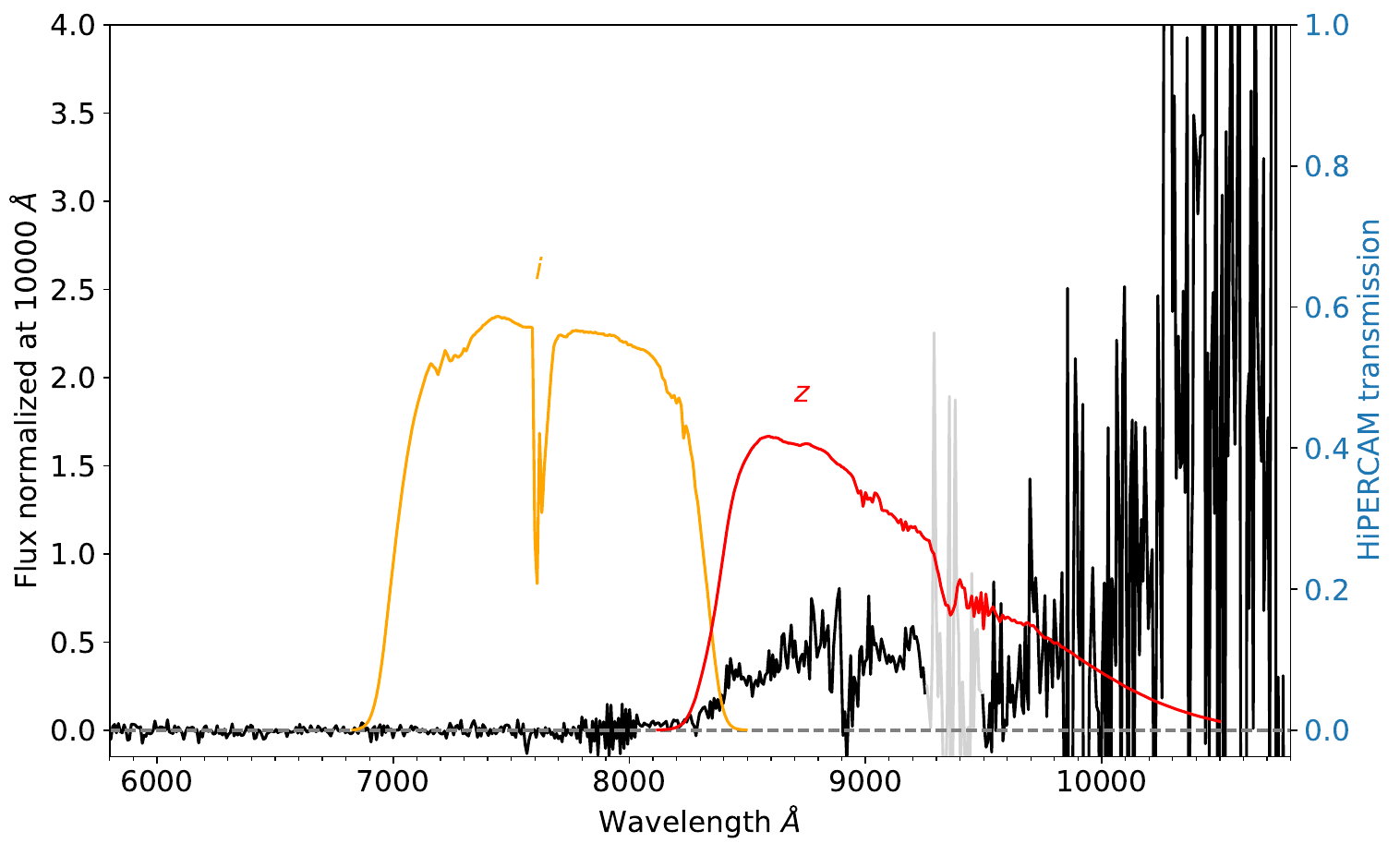}
 \caption{Calibrated and telluric corrected GTC/OSIRIS spectrum of W1738. In total, six-epoch 
spectra in 2014 and 2021 were co-added, but the part bluer than 8300 \AA \,was co-added 
using only three epochs in 2021 because in 2014 the object was contaminated by a background 
galaxy. The spectrum in light grey is dominated by telluric correction noise. Full transmission 
of the GTC/HiPERCAM $i$ and $z$ filters, including atmosphere, optics, and CCD response, are overplotted.}
 \label{original_spec}
\end{figure}

\begin{figure}
 \centering
 \includegraphics[width=9cm, angle=0]{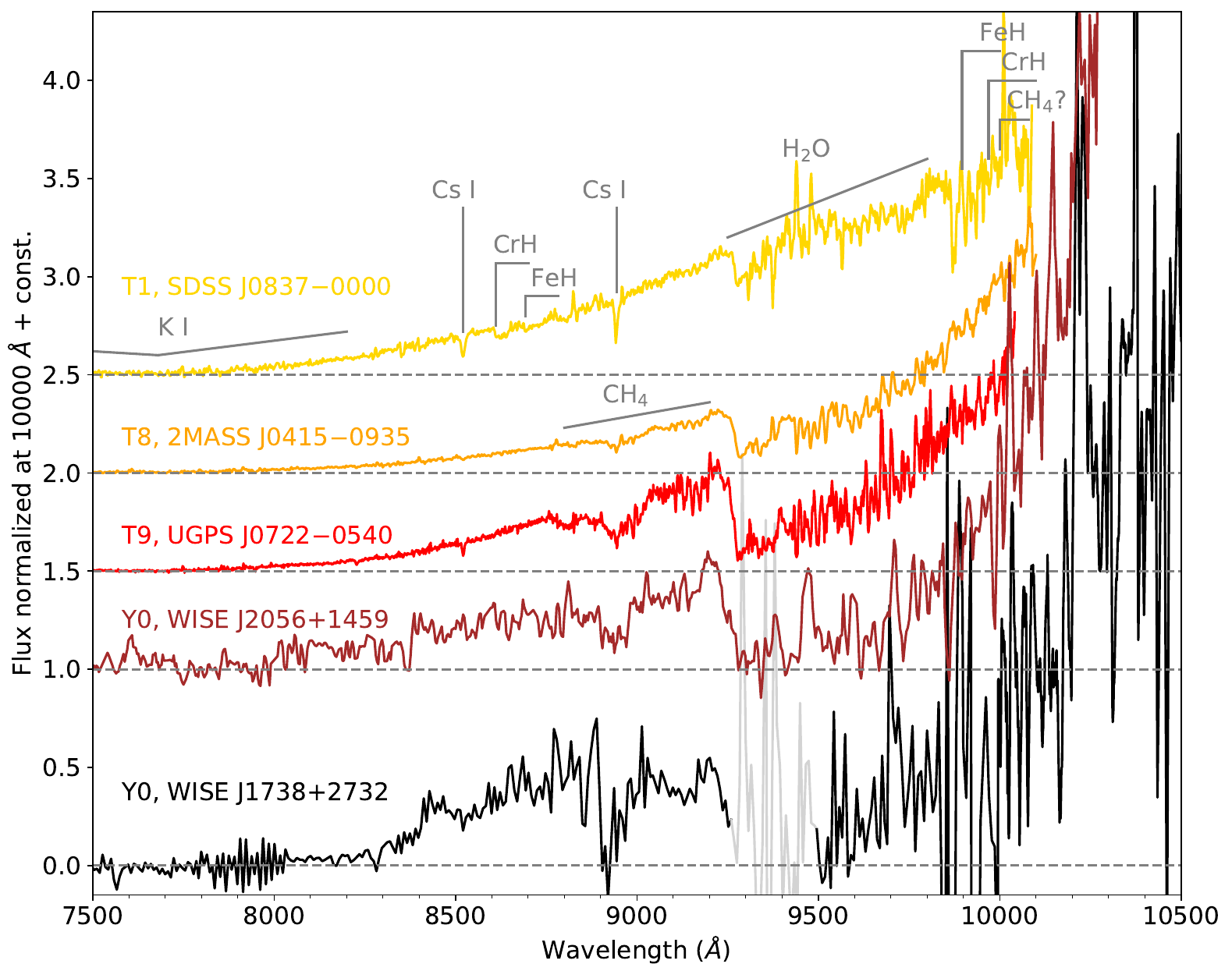}
 \caption{GTC/OSIRIS spectrum of W1738 co-added with the spectrum 
from 
\citet{leggett16} (only 
in the range of 10\,000--10\,800 \AA ) and compared with optical T1 and T8 dwarf spectra from 
\citet{burgasser13a}, an optical T9 spectrum  from \citet{leggett12a}, and a Y0 dwarf 
spectrum
from \citet{leggett13}.}
 \label{speccompare}
\end{figure}

 Figure~\ref{original_spec} shows our final GTC spectrum combining all six of the 
observing
epochs. 
The part below 8300 \AA\ was co-added using only three 
epochs
 in 2021 because in 2014 the object was contaminated by a blue background galaxy that can be seen 
in the $i$-band but not in the $z$-band. We discuss this in detail in Section~\ref{conclusions}. 
The spectrum will be made available via the GTC public archive.
The 
far-red 
optical spectrum of W1738 is shaped by molecular absorption bands and the 
extremely broad wings of the 
K I (7667/7701 \AA) resonance doublet.
Figure~\ref{speccompare} shows a comparison of our GTC spectrum with those of T and Y 
dwarfs. The wide water band (9300--9400 \AA ) can be easily recognised as the strongest
feature in the 
far-red optical spectrum of W1738. It increases in strength towards later 
T-type dwarfs and continues to get stronger in our Y dwarf spectrum. The methane feature 
(8800--9100 \AA ) is also stronger in W1738 than in the T8 standard. There is a hint of 
a detection of an absorption feature around 1 micron, possibly owing to methane 
\citep{Yurchenko17}, that has also been noted in the 
\textit{HST} spectrum of W1828  \citep{Cushing2021}, and in this paper we suggest it could be ascribed to hydrogen sulphide.   

\begin{figure}
 \centering
 \includegraphics[width=9cm, angle=0]{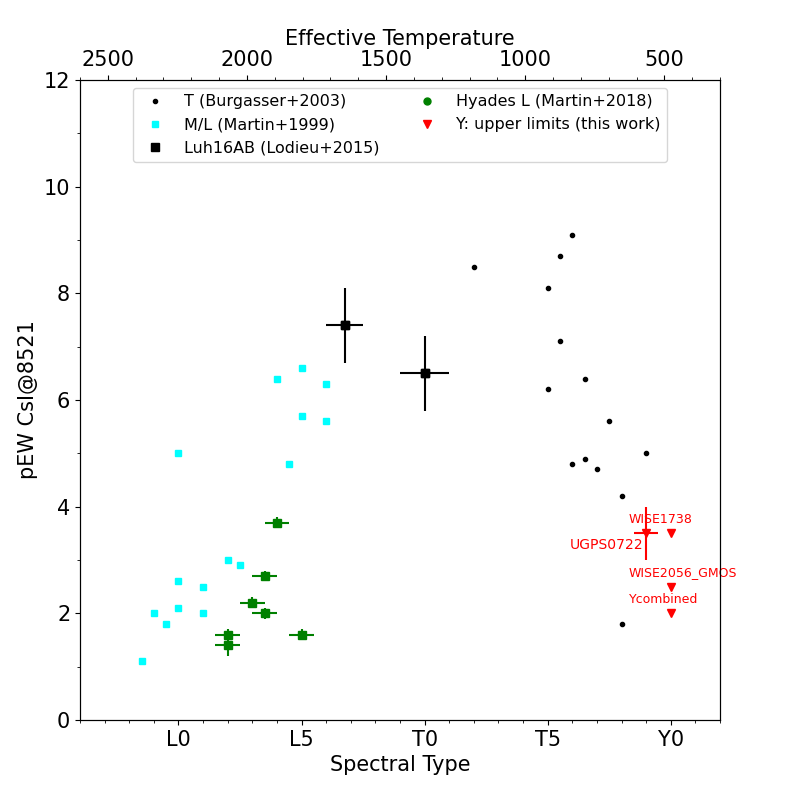}
 \caption{GTC/OSIRIS upper limit on the pEW of the Cs I resonance line (8521 \AA ) of W1738 
compared with the measurement in the T9 dwarf UGPS0722, the upper limit on the Y dwarf W2056, 
and measurements in other ultracool dwarfs from the literature. The combined upper limit for Y 
dwarfs was obtained from the sum of the W1738 and W2056 spectra.}
 \label{cs}
\end{figure}

The Cs I doublet (8521/8965 \AA ) that is prominent in L and T dwarfs (\cite{martin97b,martin99a}; 
\cite{burgasser03d}; \cite{lodieu15b,lodieu18b}) could not be detected in our GTC spectrum of W1738. 
Using the IRAF 
\texttt{splot}
task 
to estimate the 
signal-to-noise 
ratio and applying the 
equation provided in \citet{martin18a}, 
an upper limit was estimated for the pEW of the Cs I 
line at 8521 \AA. This result is shown in Figure~\ref{cs} together with measurements and 
upper limits for L and T dwarfs from the literature. W1738 follows the trend of decreasing 
Cs I strength already seen in late-T dwarfs. In the figure, additional data points for Y dwarfs 
observed with OSIRIS at GTC and available 
in
the public archive are shown, and the point 
labelled as Y combined is the upper limit of the pEW obtained by combining all the Y dwarf 
optical spectra observed with the GTC. The upper limit labelled WISE2056-GMOS was derived 
by us using the Gemini spectrum published by \citet{leggett13} and was kindly provided to us by Sandy Leggett.

%

\begin{figure}
 \centering  \includegraphics[width=9cm, angle=0]{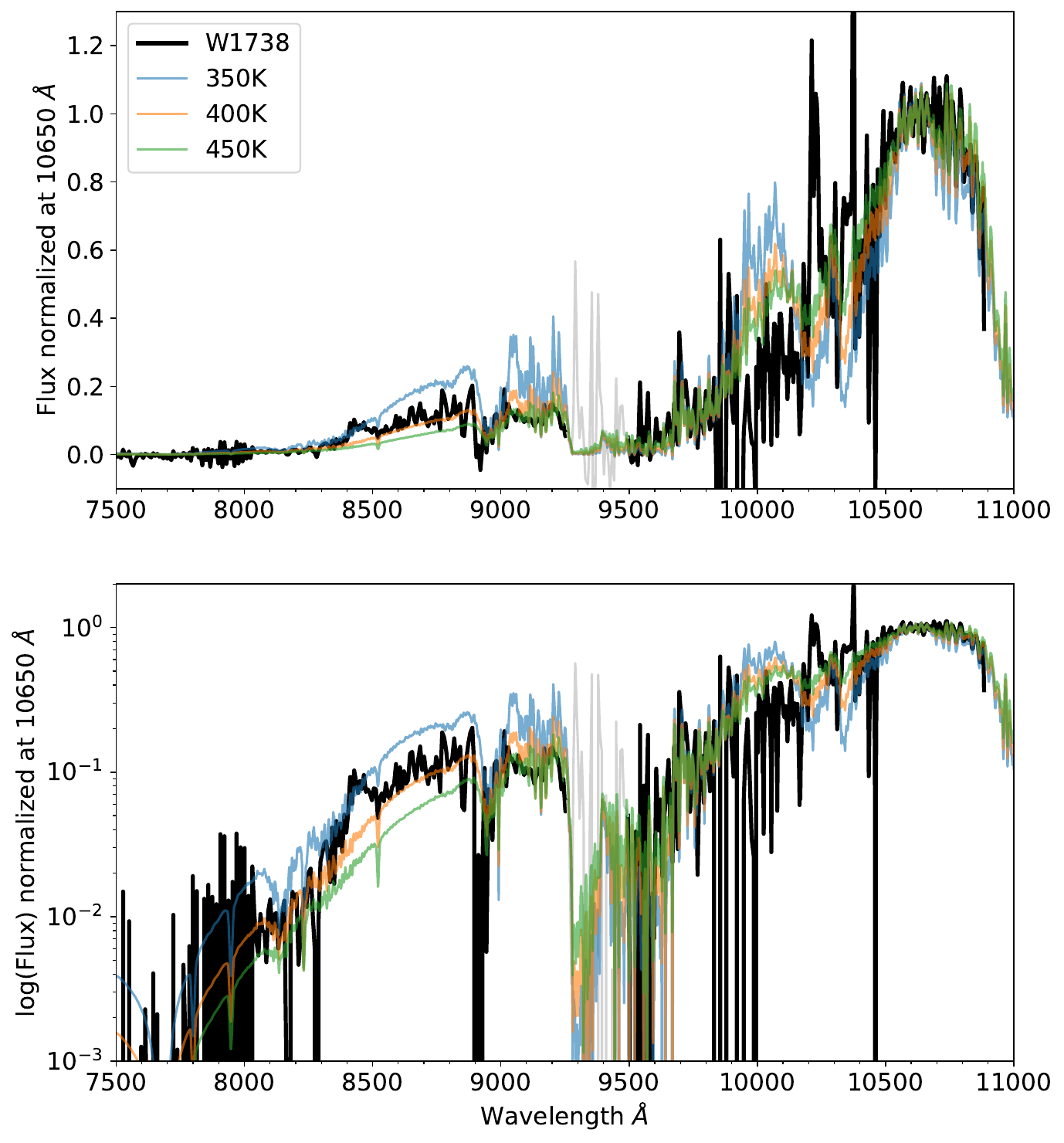}
  \caption{GTC/OSIRIS optical spectrum of W1738, co-added with the spectrum by \cite{leggett16} 
in the range 10\,000--10\,900 \AA , compared with theoretical spectra of different temperatures 
from \citet{morley14a}. The best overall agreement between the observed spectrum and the
 theoretical spectrum is for an effective temperature of 400 K. 
  The lower panel is the same as the upper panel but with 
a
log scale in the $y$-axis to 
enhance 
the low-flux region.}
 \label{specmorley}
\end{figure}

A comparison of our OSIRIS spectrum of W1738 with theoretical spectra from 
\citet{morley14a}
is shown in Figure~\ref{specmorley}. A best fit between our GTC optical spectrum and 
theoretical spectra was found for 
$T_{\rm eff} = 400$ K,
which 
is slightly lower than recent 
estimates by \citet{Kirkpatrick2021} and \citet{leggett21}, who  
obtained $T_{\rm eff} = 450$ K.
 All of these comparisons were made under the assumption of solar
metallicity and high
gravity 
because W1738 is a standard Y0 dwarf. The absolute Sloan $z$ magnitude of W1738, 23.23$\pm 0.13$ 
mag, matches a higher temperature (475 K) in the SONORA evolutionary model provided by 
\citet{marley21} despite the fact 
that
the gravity value changes from 
$\log g = 3.5$ to 5.5.
The minimisation technique to obtain the agreement between the observational and theoretical 
spectra was the same as in \citet{pavlenko02}.  
The overall agreement between the spectrum of W1738 and the synthetic spectra with $T_{\rm eff}$ 
in the range 400--450 K is fairly reasonable, although the breadth and depth of the K I 
resonance doublet is clearly overestimated in the 450 K theoretical spectrum.
It is a well-recognised challenge to model the extremely strong 
pressure-broadened 
shape of 
the K I and Na I resonance doublets in ultracool dwarfs 
\citep{marley02,burrows03b,allard05,Allard16,Pavlenko07,Phillips20},
and it continues to be a matter 
of ongoing theoretical research \citep{Allard23}. Our results for W1738 indicate that the 
theoretical treatment of K I resonance line formation in the models of Y dwarfs needs to be revised. 

\begin{figure}
 \centering  
 \includegraphics[width=9cm, angle=0]{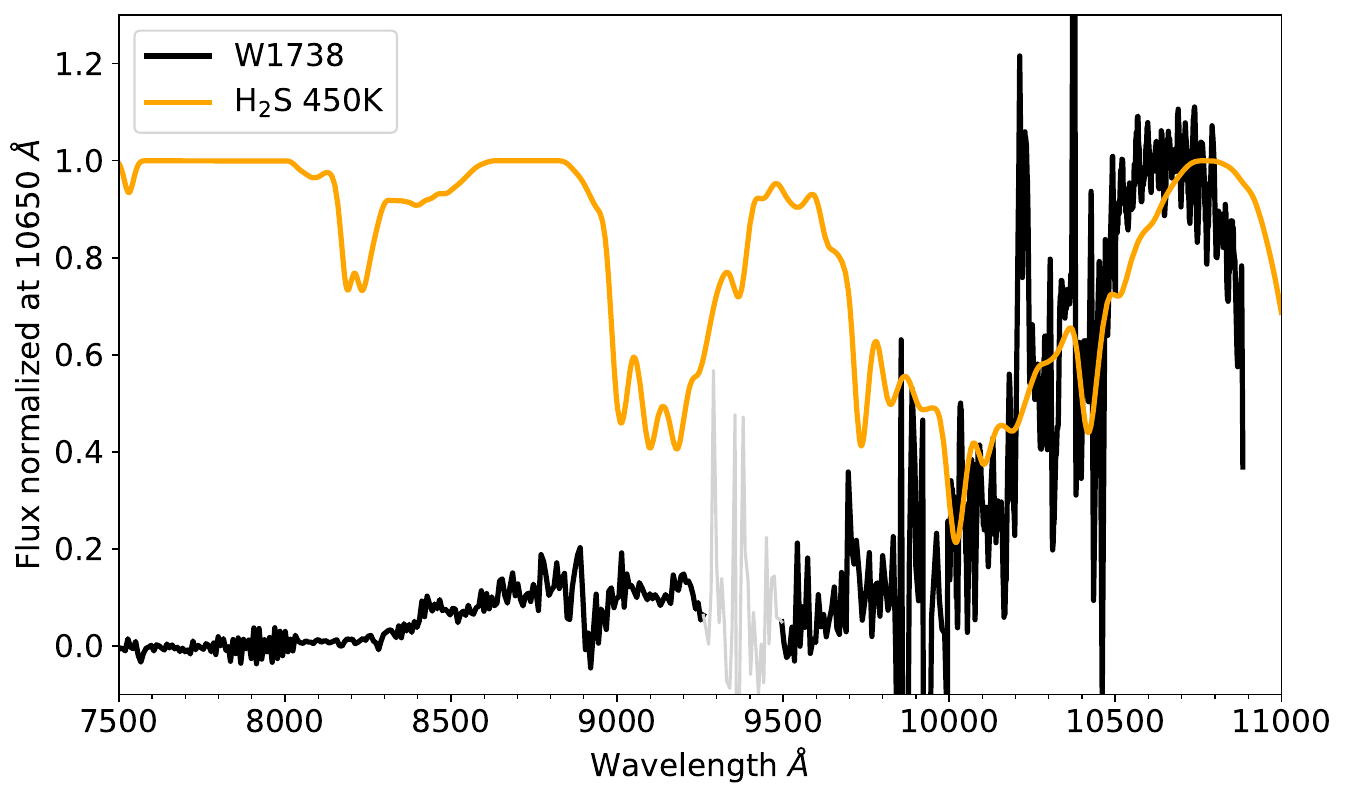}
  \caption{GTC/OSIRIS optical spectrum of W1738, co-added with the
spectrum by Leggett et al.\ (2016) in the range 10\,000–-10\,900 Å, compared with theoretical 
1D-slab far-red optical and near-infrared spectrum of H$_2$S calculated for a temperature 
of 450 K and a resolving power of 200. Several absorption features of H$_2$S can be seen that 
could correspond to those observed and 
are
still unidentified in the W1738 spectrum.}
 \label{spech2s}
\end{figure}

There
is 
a
noteworthy 
presence of 
certain
absorption features in W1738 between 8000 and 8600 \AA\
that are not seen 
in 
either
the T dwarf spectra 
or
in the synthetic spectra of \citet{morley14a}. 
Hydrogen sulphide (H$_2$S) has recently been detected in the near-infrared spectrum of a T6 dwarf 
\citep{Tannock22}. To check whether or not there could be an absorption band from H$_2$S between 
8000 and 8600 \AA, we made a 1D-slab simulation with a temperature of  450 K  and a column density 
of  $m(\mathrm{H_2S})=n(\mathrm{H_2S})\times l = 10^{30}cm^{-2}$, where 
$n(\mathrm{H_2S})$ and $l$ are the molecular density of H$_2$S and the geometrical size of the slab, respectively. 
The step in wavelength was 0.05 \AA, the resolving power was 200, and the partition function was 
$U(T) = 967.4$. The list of 
the H$_2$S lines
were taken from 
the
EXOMOL database \citep{Azzam16}.
The results are shown in Figure~\ref{spech2s}. There are absorption features of H$_2$S between 
8000 and 8600 \AA and also between 9000 and 10\,500 \AA .

\section{Discussion}
\label{discussion}
\begin{figure}
 \centering  \includegraphics[width=9cm, angle=0]{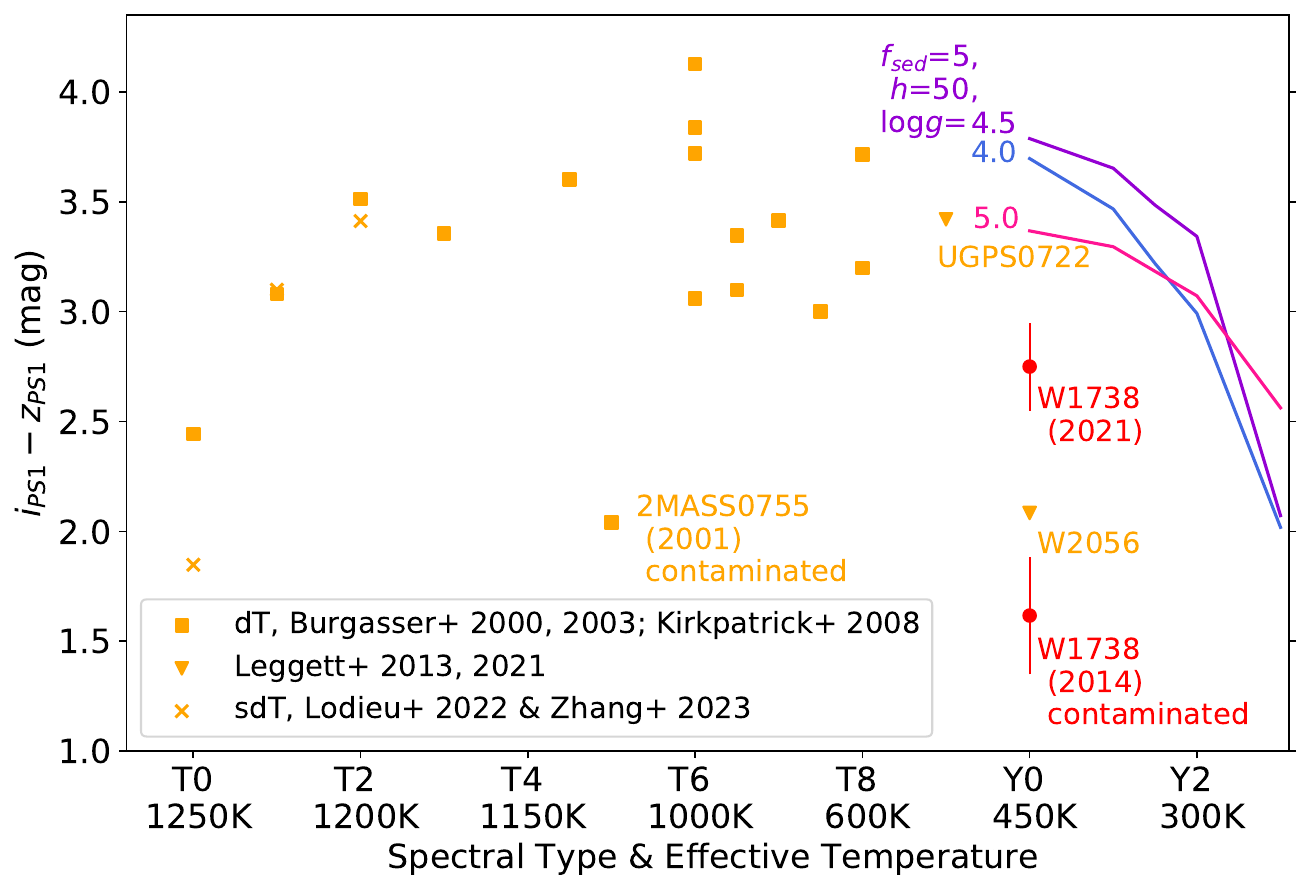}
  \caption{Predicted dependence of PS1 $i-z$ 
colours 
with spectral type of late-T and Y 
dwarfs using the observational spectra of T dwarf standards from 
\citet{burgasser00b,burgasser03b, kirkpatrick08}; T subdwarfs from 
\citet{lodieu22,zhang23}; UGPS0722 and W2056 \citep{leggett13, leggett21}; 
and the theoretical spectra of \citet{morley14a}, respectively. We adapted the 
spectral type--effective temperature relation using Table 1 in \citet{Martin2021}. 
   Three curves represent different surface 
gravities 
of the model, with fixed 
sedimentation efficiency and cloud cover. The $i-z$ 
colours 
of the Y dwarfs (W1738 and W2056) 
are bluer than these models for all the available choices of physical parameters.
 The mid-T dwarf 2MASS J07554795+2212169 and W1738 in 2014 were contaminated by two blue galaxies.} 
 \label{i-z}
\end{figure}
\begin{figure*}
 \centering  \includegraphics[width=\textwidth, angle=0]{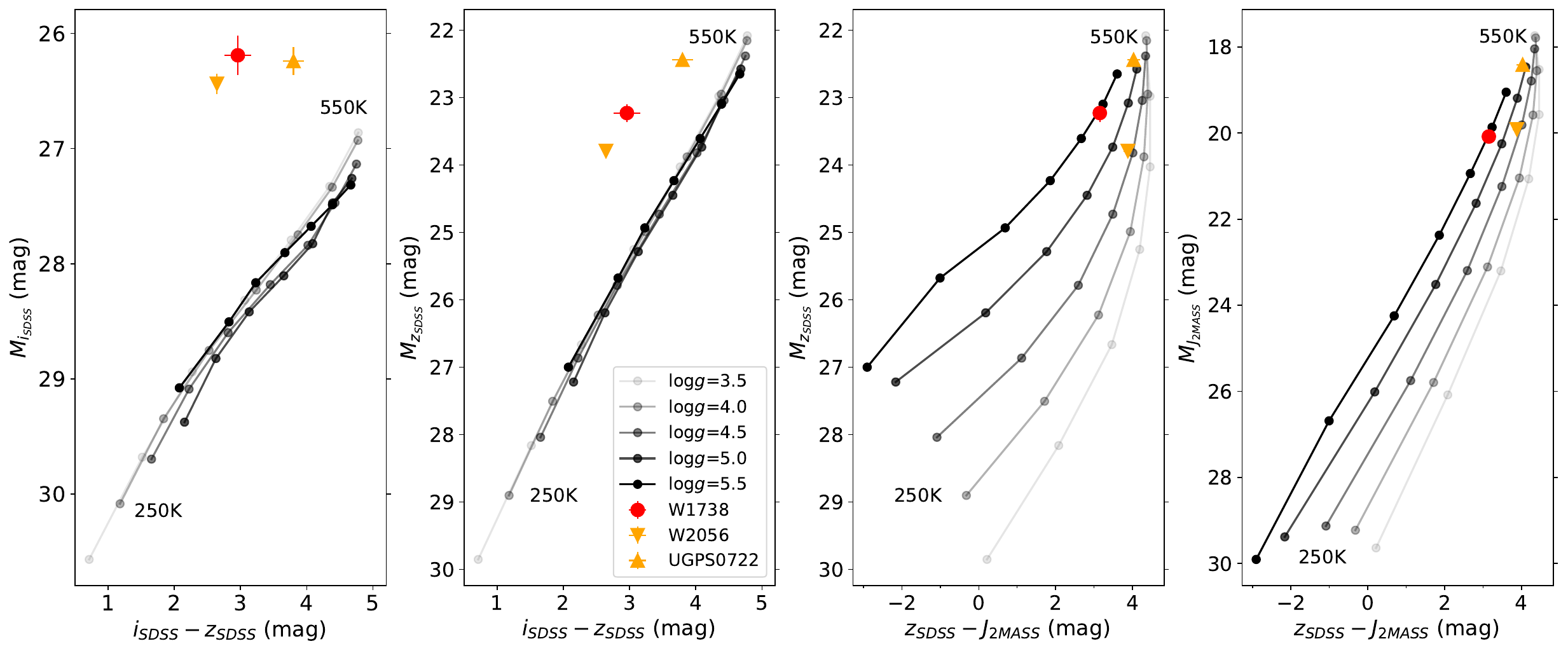}
  \caption{Hertzsprung–Russell diagrams of absolute Sloan $i$ and $z$ magnitude 
versus $i-z$ 
colour 
and absolute $z$ and $J$ magnitude versus $z-J$ 
colour. 
W1738, 
W2056, and UGPS0722 were plotted directly using their photometry, except that the 
$i$-band magnitude of W2056 is derived from its $z$-band magnitude and the $i-z$ 
colour is 
from its spectrum. 
The
SONORA solar-metallicity theoretical model from \citet{marley21} 
of different surface gravities from 3.5 to 5.5 and of effective temperatures from 
550 K to 250 K with a step of 50 K are plotted. We concluded that the abnormality 
of the blue $i-z$ 
colour 
of the Y dwarfs in the Figure~\ref{i-z} is caused by its excess flux 
in the $i$-band.} 
 \label{i-z vs M}
\end{figure*}

 \begin{figure}
 \centering
 \includegraphics[width=9cm, angle=0]{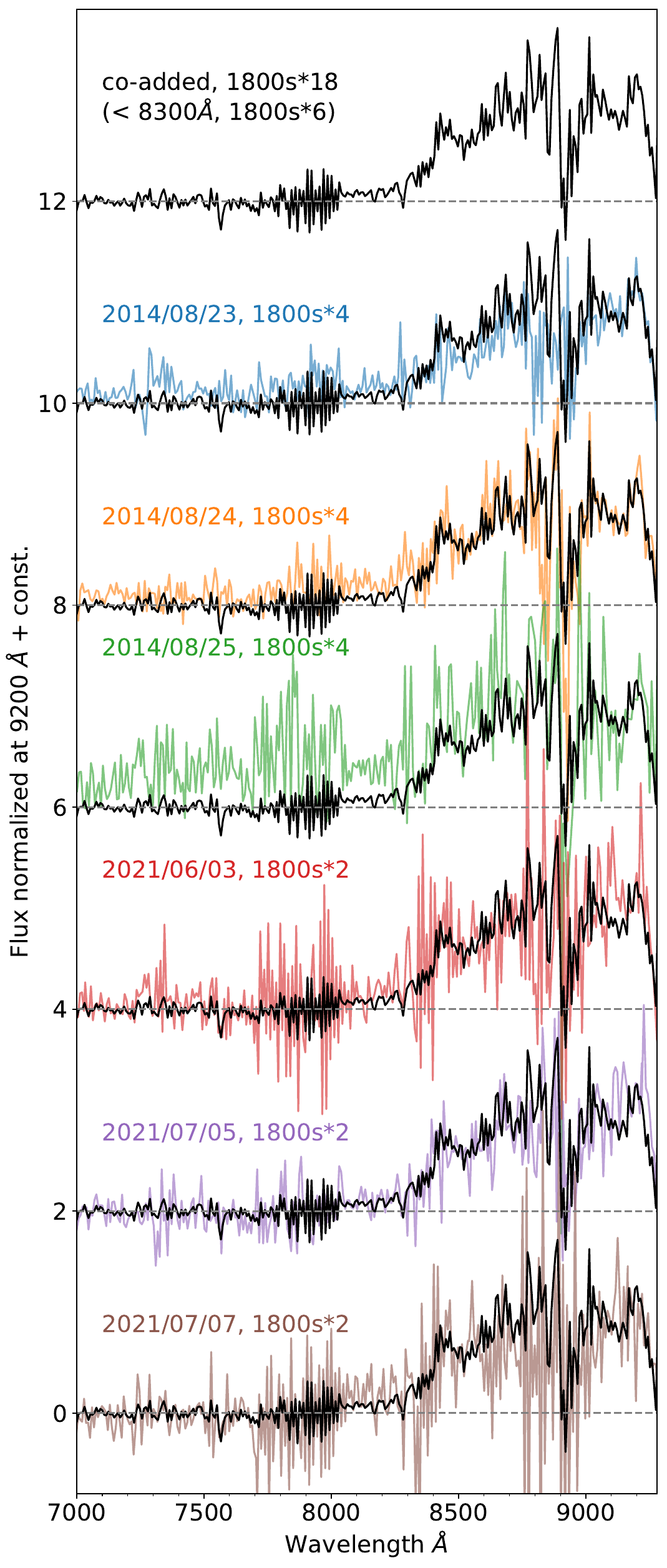}
  \caption{Zoom-in of the GTC/OSIRIS optical spectra of W1738 in the region dominated 
by methane (and possibly hydrogen sulphide) and relatively free from telluric 
absorption. The spectra obtained at six different epochs (labelled) are compared 
with the co-added spectrum (solid black, and the part below 8300 \AA\ was co-added 
using only the spectra in 2021). In 2014, W1738 had more flux in the $i$-band than 
in 2021 because the Y dwarf passed very close to a background galaxy that contaminated 
blue part of the spectrum.}\label{specplot_epoch}
\end{figure}

In Figure~\ref{i-z}, we show the mean values of the $i-z$ 
colours  
that we obtained from 
observational spectra for T and Y dwarfs and from theoretical spectra for Y dwarfs 
and the $i-z$ 
colour 
derived from our GTC observations of W1738. The $i-z$ 
colours 
of the Y dwarfs are bluer than those of mid- and late-T dwarfs. 

We synthesised the PS1 photometry of W1738 from multiple-epoch spectra. The $i-z$ 
colour 
is 1.62 mag for the co-added spectrum in 2014. We could not synthesize $i$-band photometry 
for the co-added spectrum in 2021 because of the low signal-to-noise ratio.
Instead, we used the HiPERCAM photometry 
in 2021 and then applied a correction for the filter profiles (the
main difference is the two $z$ 
filters).  The error bar of W1738 in 2021 was estimated based on photometric errors, and that 
of 2014 was scaled based on the difference of the efficiency between GTC/OSIRIS and GTC/HiPERCAM, 
on-source exposure time, and magnitudes in 
the
$i$- and $z$-bands in two epochs. As a result, in 
2021 $i-z = 2.75$ mag, indicating a 1.13 mag change in $i-z$ 
colour between the 2014 and 2021 
epochs of observation, with a significance of 
2.5$\sigma$. 
As explained below, this is 
probably
the result of contamination by a galaxy in the 2014 spectrum. 

The 
blueing
of the $i-z$ 
colours 
in the Y dwarfs is predicted by the models because of diminished 
absorption of the huge Na I and K I resonance doublets caused by grain sedimentation \citet{marley02}. 
Alkali elements have low condensation temperatures, and in the atmospheres of Y dwarfs, they become 
locked into molecules, such as 
KCl  \cite{marley02},  
which 
can condense out from the 
gas phase into solid dust grains.
A confirmation of the weakening of the resonance lines of alkali elements in the atmospheres of Y 
dwarfs comes from the non-detection of the Cs I resonance features in our optical spectrum of W1738. 
An independent confirmation also comes from the fact that the disagreement between the model predictions and 
observations occurs when using the $i$-band, but not when using 
$z$ and $J$, as shown in  
Figure~\ref{i-z vs M}. 
 
The information obtained from the GTC spectrum of W1738 presented in this 
paper 
shows that optical 
ground-based spectroscopy can complement space-based infrared spectroscopy. This is particularly 
timely, as detailed spectroscopic studies of Y dwarfs are starting to be carried out with JWST 
\citep{Beiler2023}.

Our upper limits on the optical emission of Y dwarfs ($u$-,$g$-, $r$- bands) are the deepest ever published 
for these objects and set constraints on exotic sources of non-thermal optical light emission that 
could 
arise 
from phenomena such as aurorae, bioluminescence, and lightning. 
Y dwarfs are expected to be extremely faint in the $g$ filter. Using the models of 
\citep{morley14a}, we estimate the predicted $g-z$ 
colour 
from photospheric emission to be $\ge 6$ for Y dwarfs. 
Our $g-z \ge 3.57$ 
colour 
limit for W1738 sets a 
constraint on the presence of non-photospheric 
visible light emission in this object. It 
might 
be worthwhile to reach deeper than our GTC limit 
in order to detect the photospheric emission contribution in the $g$-band.

After 
W1738 was observed at multiple epochs with the long-slit mode of OSIRIS at 
the GTC, 
the spectra were compared to check for the possibility of variability. This object has 
been reported to be photometrically variable at 1 and 4 microns \citep{leggett16}. 
As shown in Figure~\ref{specplot_epoch}, in 2014 W1738 showed more flux in the $i$-band, 
and in 2021 it 
had
more flux around 9100 \AA. The $i-z$ 
colour 
in 2014 indeed was 1.13 mag 
bluer than that in 2021. However, this bluer 
colour 
is spurious, owing to contamination by 
a galaxy at 17h38m35.69s$+$427$^\circ$32'57.8" in 2014. The galaxy is not visible in the deep 
$z$-band images nor
in 
the GTC/OSIRIS data obtained in 2012 by \cite{lodieu13a}
or
the GTC/HiPERCAM data obtained by us in 2021, but it is clearly visible in our
 GTC/HiPERCAM $u$-, $g$-, $r$-, and $i$-bands. We also checked the anomalous blue optical 
colour 
of 2MASS J07554795+2212169 for its spectral class (T5--T6) in Figure~\ref{i-z}, 
and we found that the optical spectrum was also 
probably
contaminated by a blue galaxy when 
its spectrum was taken in 2001. The issue of line-of-sight coincidences between extremely 
faint ultracool dwarfs and galaxies deserves further investigation, and we plan to study it 
in the near future. 
The upper bound of peak-to-peak variation 
at the 
day and month level from the synthesised 
photometry is of 0.53 mag for the $z$-band, 
including any possible instrumental variations or 
changes in weather conditions.

%

\section{Final remarks}
\label{conclusions}

In this 
paper, 
we have presented optical imaging observations of five Y dwarfs and a long-slit 
far-red spectrum for one of the dwarfs obtained at the GTC. 
One of our targets, 
W1738, was clearly detected in the $i$-band, the first such 
detection of a Y dwarf. We obtained photometric measurements of W1738 in the $z$- and $i$-bands, 
a low-resolution 
far-red 
optical spectrum, and upper limits to the emission 
in the 
$r$-, $g$-, and $u$-bands.
The $i-z$ and $z-J$ 
colours 
of W1738 are bluer than 
for
most T dwarfs. The $z-J$ 
colour 
of W1828 is even bluer than that of W1738. 
The most likely explanation for the 
colours 
of W1738 and W1828 is that they are relatively 
blue because of the weakening of the very broad K I resonance doublet. 
The K I resonance feature of the synthetic spectra is too strong and makes the predicted 
$i$-band flux and $i-z$ 
colour 
fainter and redder, respectively, than the observations of 
Y dwarfs. The theoretical  broadening and depth of the alkali lines needs to be revised. 
The presence of methane and water was detected in the spectrum of W1738 but not 
in
that of 
Cs I. This is consistent with the weakening of Cs I features observed in other Y dwarfs. 
There are absorption features (8000--8600 \AA ) in our GTC spectrum of W1738 that are not 
seen in the 
far-red 
spectra of T dwarfs. The 1D-slab simulations suggest that H$_2$S has 
absorption features in this spectral range; 
hence, hydrogen sulphide could be responsible 
for those features in our optical spectrum of W1738. We found a change in the $i-z$ 
colour 
of W1738 between the 2014 and 2021 OSIRIS observations that we attribute to contamination 
of the $i$-band long-slit spectrum by a galaxy in 2014. Care must be exercised because 
line-of-sight chance alignments with background galaxies could be quite common. We have 
shown that the
$i$-band 
flux of Y dwarfs is brighter than expected based on theory, and we note 
that this finding could help detect this type of object in upcoming large-area surveys such as the \textit{Euclid} space mission and 
the Vera C. Rubin Legacy Survey of Space and Time, which  
will obtain deep observations at optical wavelengths. The number of Y dwarfs that will be detected 
by
\textit{Euclid} at near-infrared wavelengths is not expected to be large \citep{Martin2021,Solano2021}, 
and in the optical, this number has not been estimated.

%
%
\begin{acknowledgements}
The anonymous referee is gratefully acknowledged for three insightful reports that 
greatly improved the contents and presentation of this paper. Funding for this work was 
provided by the European Union (ERC, SUBSTELLAR, project number 101054354) and the Agencia 
Estatal de Investigación del Ministerio de Ciencia e Innovación (AEI-MCINN) under grant  
PID2019-109522GB-C53\@. YP's work has been carried out within the framework of the MSCA4Ukraine program, 
project number 1.4-UKR-1233448-MSCA4Ukraine.
This paper is based on observations made with the Gran Telescopio Canarias (GTC), installed 
at the Spanish Observatorio del Roque de los Muchachos of the Instituto de Astrofísica de 
Canarias, on the island of La Palma. This work is (partly) based on data obtained with the 
OSIRIS
instrument, 
built by a Consortium led by the Instituto de Astrofísica de Canarias in 
collaboration with the Instituto de Astronomía of the Universidad Autónoma de México. OSIRIS 
was funded by GRANTECAN and the National Plan of Astronomy and Astrophysics of the Spanish Government. 
This research has made use of the Simbad and Vizier databases, operated
at the centre de Donn\'ees Astronomiques de Strasbourg (CDS), and
of NASA's Astrophysics Data System Bibliographic Services (ADS). 
The Pan-STARRS1 Surveys (PS1) and the PS1 public science archive have been made possible 
through contributions by the Institute for Astronomy, the University of Hawaii, the Pan-STARRS 
Project Office, the Max-Planck Society and its participating institutes, the Max Planck Institute 
for Astronomy, Heidelberg, and the Max Planck Institute for Extraterrestrial Physics, Garching, 
The Johns Hopkins University, Durham University, the University of Edinburgh, the Queen's 
University Belfast, the Harvard-Smithsonian Center for Astrophysics, the Las Cumbres Observatory
Global Telescope Network Incorporated, the National Central University of Taiwan, the Space 
Telescope Science Institute, the National Aeronautics and Space Administration under 
Grant No.\ NNX08AR22G issued through the Planetary Science Division of the NASA Science 
Mission Directorate, the National Science Foundation Grant No. AST-1238877, the University 
of Maryland, Eotvos Lorand University (ELTE), the Los Alamos National Laboratory, and the 
Gordon and Betty Moore Foundation. 
\end{acknowledgements}
%

%
%
\bibliographystyle{aa}
\bibliography{biblio_new5}

\end{document}